\documentclass[twocolumn,floatfix,aps,pre,amsmath,showpacs]{revtex4}

\usepackage{graphicx}

\begin{document}
\title{Democratic particle motion for meta-basin transitions in simple glass-formers}
\author{ G. A. Appignanesi$^\dag$, J. A. Rodr\'{\i}guez Fris$^\dag$, R. A. Montani$^\dag$ 
and W. Kob$^\ddag$}
\affiliation{
$^\dag$  Laboratorio de Fisicoqu\'{\i}mica, Departamento de Qu\'{\i}mica, Universidad Nacional del Sur, Av. Alem 1253, 
8000 Bah\'{\i}a  Blanca, Argentina.\\
$^\ddag$ Laboratoire des Collo\"ides, Verres et Nanomat\'eriaux, Universit\'e Montpellier 2, 
34095 Montpellier, France.\\
}
\date{\today}

\begin{abstract}

We use molecular dynamics computer simulations to investigate the local
motion of the particles in a supercooled simple liquid. Using the concept
of the distance matrix we find that the $\alpha-$relaxation corresponds
to a small number of crossings from one meta-basin to a neighboring
one. Each crossing is very rapid and involves the collective motion of
$O(40)$ particles that form a relatively compact cluster, whereas
string-like motions seem not to be relevant for these transitions. These
compact clusters are thus candidates for the cooperatively rearranging
regions proposed long times ago by Adam and Gibbs.

\end{abstract}


\maketitle

In recent years significant progress has been made in our understanding
of the relaxation dynamics of glass-forming liquids at intermediate and
low temperatures. Sophisticated experiments and computer simulations
have identified many of the salient features of this dynamics, and
theoretical approaches have helped to rationalize them, at least to
some extend~\cite{ediger_96,gotze_99,debenedetti_01,ngai_02}. Despite
this progress, many of the most elementary questions have
not been answered so far and among them is the nature of the
motion of the particles in the $\alpha-$relaxation regime at
low temperatures. Experiments and simulations have demonstrated
that this dynamics is quite heterogeneous and therefore can
be used to explain the observed stretching of the time correlation
functions~\cite{schmidt_91,butler_91,cicerone_95,kob_97,kegel_00,weeks_00,richert_02}.
This heterogeneous dynamics has been shown to be related to cooperative
motion in which a small number of particles (a few percent)
undergo a collective relaxation dynamics in that they move, often
in a string-like fashion, by a distance that is comparable to the
one between neighboring particles~\cite{donati_98,kegel_00,weeks_00}.
Since a qualitatively similar heterogeneous dynamics has also been found
in simple lattice models that show a glassy dynamics and for which it
is well known that the $\alpha-$relaxation is intimately connected
to the dynamical heterogeneities (DH), it has been proposed that
these DH are crucial for the relaxation dynamics of all glass-forming
systems~\cite{garrahan_02}. However, since in these lattice models all
elastic or quasi-elastic effects are completely neglected, it is not at
all clear whether or not the DH are indeed the only relevant mechanism
for the relaxation.

Another approach to describe the relaxation dynamics on the time scale
of the $\alpha-$relaxation is by means of the so-called potential energy
landscape (PEL)~\cite{angell_91,debenedetti_01b,monthus_96} (or more
precisely the free energy landscape) and hence to describe the dynamics
of the system by considering its trajectory in configuration space. This
PEL is rugged due to the presence of barriers in the free energy and
hence at low temperatures the resulting dynamics will be slow. Using the
concept of the inherent structures, evidence has been given that (roughly
speaking) the motion of the system in the PEL can be decomposed into two
types of movements~\cite{doliwa_03,vogel_04}: In the first type the system
explores some minima which are locally connected to each other and are not
separated by a significant barrier. Therefore such a collection of minima
is called ``meta-basin'' (MB) and its exploration corresponds to the
$\beta-$relaxation. As a second type of motion, the system overcomes the
barriers that surround a MB and enters a new MB, a motion that has been
believed to correspond to the $\alpha-$relaxation~\cite{debenedetti_01}
or to the elementary events of the $\alpha-$relaxation~\cite{vogel_04}.

Although this picture for the motion of the system within the PEL is
certainly appealing from a qualitative point of view, it does not
give any insight on the nature of the dynamics of the particles on
the microscopic level. Furthermore it remains unclear whether the DH
mentioned above have anything to do with the barrier-crossing of the
system moving in the PEL. The goal of the present work is therefore to
clarify this issue and thus to make an advancement in our understanding
of the relaxation dynamics of supercooled liquids. To this aim we have
done molecular dynamics computer simulations of a simple glass-former
in order to identify the presence of the MB and to investigate the
nature of the motion of the particles during the transition from one
MB to another. The so obtained results can then be compared with the
heterogeneous motion of the particles in order to see to what extend
the two motions are related to each other.

The system considered is a binary mixture of Lennard-Jones (LJ) particles.
In previous investigations it has been shown that this system shows
many features of glass-forming liquids and can thus serve as a simple
model for such liquids~\cite{kob_95}. The interaction between two atoms
of type $A$ (80\%) and $B$ (20\%) is given by $V_{\alpha \beta}(r)
= 4 \epsilon_{\alpha \beta} \{ (\sigma_{\alpha \beta}/r)^{12} -
(\sigma_{\alpha \beta}/r)^{6} \}$, where $\alpha, \beta \in \{A,B\}$. The
LJ parameters used are $\epsilon_{AA} = 1.0$, $\sigma_{AA} = 1.0$,
$\epsilon_{AB} = 1.5$, $\sigma_{AB} = 0.8$, $\epsilon_{BB} = 0.5$,
and $\sigma_{BB} = 0.88$. These interactions have been truncated and
shifted at $r_{\rm cutoff}=2.5\sigma_{\alpha\beta}$. In the following
we will use $\sigma_{AA}$ and $\epsilon_{AA}$ as units of length and
energy, respectively, and measure time in units of $(m \sigma_{AA}^{2}
/ 48 \epsilon_{AA})^{1/2}$. The equations of motion were solved for the
NVE ensemble at a particle density of 1.2, using the velocity form of
the Verlet algorithm with a time step of 0.02. All the presented results
correspond to the situation in equilibrium.

In order to identify the MBs we define the following ``distance matrix''
(DM)~\cite{ohmine_95}:
\vspace*{-4mm}

\begin{equation}
\Delta^2(t',t'') = \frac{1}{N}\sum_{i=1}^N |{\bf r}_i(t')-{\bf r}_i(t'')|^2
\quad ,
\label{eq1}
\end{equation}

\noindent
where ${\bf r}_i(t)$ is the position of particle $i$ at time $t$.  Thus
$\Delta^2(t',t'')$ gives the system averaged squared displacement
of a particle in the time interval that starts at $t'$ and ends at
$t''$. Note that the time average of $\Delta^2(t',t'+\theta)$
over $t'$ gives the $r-$average of $G_s(r,\theta)$, the self-part of
the van Hove correlation function for time displacement $\theta$. The
same is true if one averages over a very large system. Since we are
interested in {\it individual} MB-MB transitions, one has to avoid that
the presence of several independent local rearrangements, that will
occur in a large system almost simultaneously, obscures the analysis of
the individual event. Therefore we have considered a rather small system
of 150 particles, the smallest possible system that does not affect the
interactions, i.e. the box size was two times $r_{\rm cutoff}$ for the
$A-A$ interaction~\cite{footnote1}.

\begin{figure}[tb]
\includegraphics[width=0.9\linewidth]{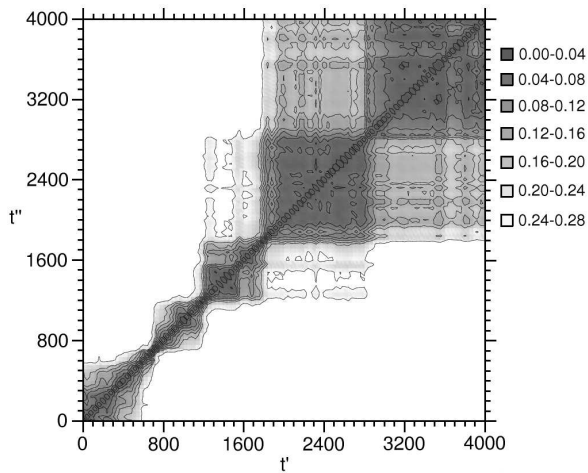}
\caption{
Distance matrix $\Delta^2(t',t'')$ of the system for $T=0.50$. 
The gray level correspond to values of $\Delta^2(t',t'')$ that are
given to the right of the figure.}
\vspace*{-3mm}

\label{fig1}
\end{figure}

\begin{figure}[tb]
\vspace*{4mm}

\includegraphics[width=1.0\linewidth]{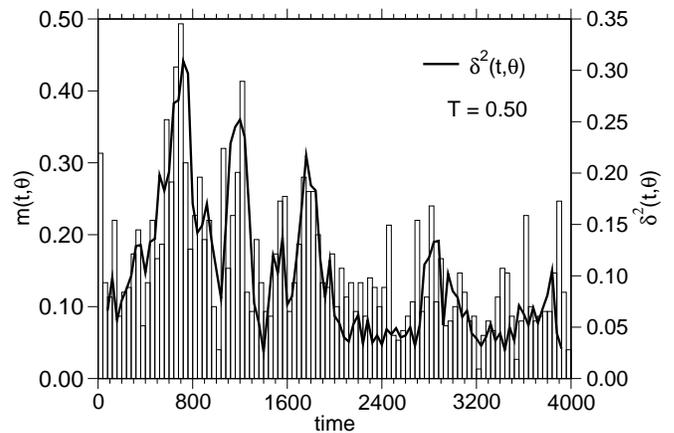}
\caption{
Solid line (right scale): Averaged squared displacement $\delta^2(t,\theta)$ for
the trajectory given in Fig.~\ref{fig1}. The value of $\theta$ is 160.
Vertical bars (left scale): The function $m(t,\theta)$ which gives the fraction of
mobile particles, i.e. particles that moved more than the threshold value
$r_{\rm th}= 0.3$ in the time interval $[t,t+\theta]$, using $\theta=40$.}
\label{fig2}
\end{figure}

Figure~\ref{fig1} shows a typical graph of the DM at $T=0.50$ as
a function of the two time arguments $t'$ and $t''$, with darker areas
corresponding to configurations that have a smaller distance. At this
temperature the dynamics is already slow with an $\alpha-$relaxation time
$\tau$ of the order of $4\cdot 10^3$~\cite{kob_95}. ($\tau$ can, e.g. be defined
by requiring that the self intermediate scattering function $F_s(q,t)$ has
decayed to 10\% of its initial value.) From this figure we see immediately
that the dynamics of the system is quite heterogeneous {\it in time} in
that it stays for a significant time relatively close to one region in
configuration space, dark square-like regions, before it finds a pathway
to a new region. Thus this is clear evidence that the system explores the
present MB before it moves on to a neighboring one. At this temperature
the typical sojourn time within one MB is around 300-800 time units,
which is around 10\% of $\tau$. Thus this sojourn time corresponds to the
time scale of $t^*$, the time that previous investigations have shown to
be relevant for the dynamical heterogeneities in the system and which is
defined as the time at which one observes the maximum in the non-gaussian
parameter $\alpha_2(t)$~\cite{kob_97,appignanesi_04}, and which at this
temperature is $400$ time units~\cite{kob_95}. Hence we can conclude
that the MB-MB transitions are relevant for the DH whereas it takes about
5-10 such transitions in order to make an $\alpha-$relaxation. We also
point out that from Fig.~\ref{fig1} it becomes evident that the time
for a MB-MB transition is quite short, on the order of 100~time units,
which thus corresponds to about 20\% of $t^*$. Thus the transition is
significantly faster than the $\alpha-$relaxation times $\tau$.

\begin{figure}[tb]
\includegraphics[width=0.9\linewidth]{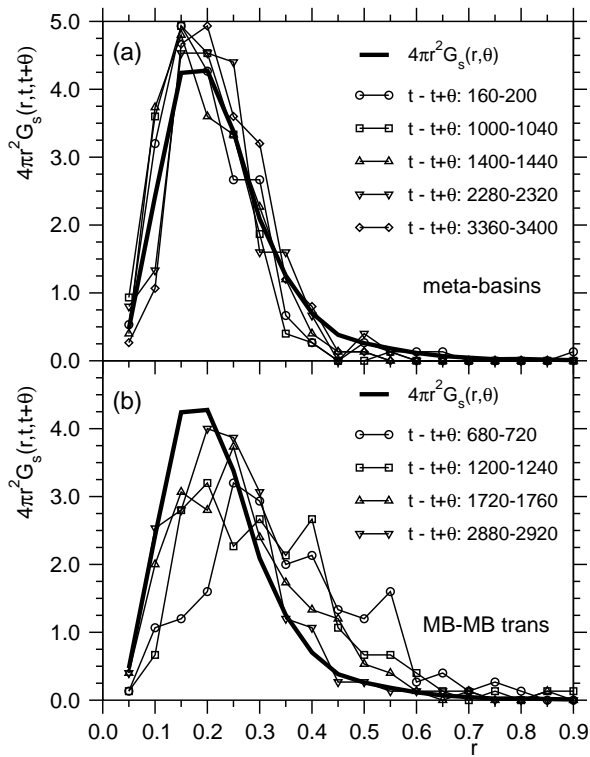}
\caption{
The distribution function $\widehat{G}_s(r,t,t+\theta)$ for different
values of $t$ (curves with symbols). The value of $\theta$ is 40. The
bold curve is $G_s(r,\theta)$, the self part of the van Hove function. a) Values
of $t$ in which the system is inside a MB. b) Values of $t$ for which the system
is about to leave a MB.}
\label{fig3}
\end{figure}

In Fig.~\ref{fig2} we show for the same run and time interval
$\delta^2(t,\theta)$, the (particle) averaged squared displacement (ASD)
of the particles within a time interval $\theta$ (solid curve). This
function is defined as
\vspace*{-4mm}

\begin{eqnarray}
\delta^2(t,\theta)  & = & \Delta^2(t-\theta/2,t+\theta/2) \\
 & = & 
\frac{1}{N}\sum_{i=1}^N |{\bf r}_i(t-\theta/2)-{\bf r}_i(t+\theta/2)|^2\quad .
\label{eq2}
\end{eqnarray}

\noindent
Thus $\delta^2(t,\theta)$ is $\Delta^2(t',t'')$ measured along the
diagonal $t''=t'+\theta$ and hence the average of this quantity over
different start times $t$ gives the usual mean-squared displacement
for time lag $\theta$.  A comparison of this ASD with Fig.~\ref{fig1}
shows that $\delta^2(t,\theta)$ is showing pronounced peaks exactly then
when the system leaves a MB. Thus we see that changing the MB is indeed
associated with a rapid motion as measured in the ASD.

In order to understand the nature of the motion of the particles when the
system leaves a MB, we have calculated $\widehat{G}_s(r,t,t+\theta)$,
the distribution of the displacement $r$ of a particle for a
given time difference $\theta=40$. (Note that the average of
$\widehat{G}_s(r,t,t+\theta)$ over $t$ gives $G_s(r,\theta)$.) This
distribution is shown in Fig.~\ref{fig3}a for starting values $t$
that correspond to times at which the ASD shows a plateau, i.e. when
the system explores a MB. Also included in the graph is the self
part of the van Hove function, $G_s(r,\theta)$, and we see that
the distributions $\widehat{G}_s(r,t,t+\theta)$ are narrower and
more peaked than $G_s(r,\theta)$ thus showing that in a MB the system
moves more slowly than on average. In Fig.~\ref{fig3}b we show the same
distributions but now for times in which the system is about to leave a
MB (compare the values of $t$ with Fig.~\ref{fig1}). For these values
of $t$ the distributions are displaced to the right with respect to
$G_s(r,\theta)$, showing that in this time regime the motion of the
system is faster than on average. Most noteworthy is the observation
that this shift is relatively uniform, i.e. a substantial part of the
particles moves quicker than on average.  Thus we can conclude that the
rapid increase of the ASD is {\it not} due to the presence of a {\it
few} fast moving particles, but instead to a ``democratic'' movement
of {\it many} particles, in contrast to the results for cooperative
motion on the time scale of $t^*$ which has been documented in earlier
work~\cite{kegel_00,weeks_00,donati_98}. Thus this movement is very
different in nature from the ``string-like'' motion found in the context
of the dynamical heterogeneities~\cite{donati_98,appignanesi_04}.

To demonstrate that the number of particles that participate at
this democratic motion is indeed substantial and strongly correlated
with a strong increase in the ASD, we have defined as ``mobile'' all
those particles that in the time interval $\theta=40$ have moved more
than $r_{\rm th}=0.3$, and denote the fraction of such particles by
$m(t,\theta)$~\cite{footnote2}.  In Fig.~\ref{fig2} we have included the
fraction of mobile particles as a function of time (vertical bars) and
a comparison of this data with the ASD in the same graph shows that the
fraction of mobile particles is indeed large whenever the ASD increases
rapidly. This fraction is on the order of 30\% of the particles and
thus significantly larger than one would expect from $G_s(r,\theta)$
if one integrates this distribution from $r_{\rm th}$ to infinity and
which gives 0.09.

\begin{figure}[tb]
\includegraphics[width=0.80\linewidth]{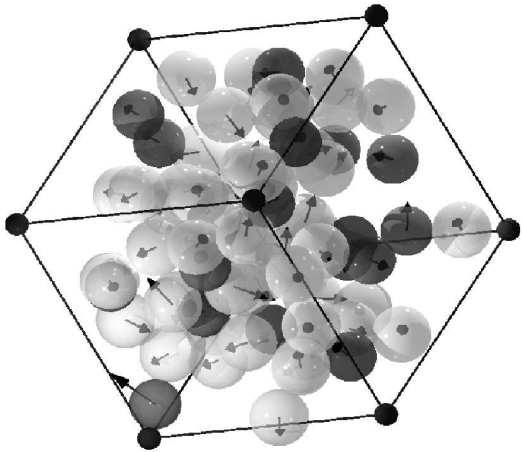}
\caption{
Configuration snapshot of mobile particles occurring in the MB-MB
transition $t=680 \to t=720$. The spheres (light and dark for the $A$
and $B$ particles, respectively) give the location of the particles
before the rearrangement and the arrows point to their position after
the transition. The black spheres give the location of the corner of
the simulation box.}
\label{fig4}
\end{figure}

In order to give an idea on the nature of the motion of the mobile
particles during a MB-MB transition, we show in Fig.~\ref{fig4} a typical
configuration of mobile particles before such a transition event and attach to each
particle an arrow which points to the location of the particle after the
event, i.e. a time $\theta=40$ later. From this graph we recognize that
the MB-MB transitions correspond to a movement in which the particles
form a relatively compact cluster. Thus this is in contrast to the
type of motion found in the context of the DH in which the particles
form string-like objects~\cite{donati_98,appignanesi_04b}. These
compact regions have, at the temperature considered, around 30-60
particles and can be considered as potential candidates for the
cooperatively rearranging regions (CRR) proposed long time ago by
Adam and Gibbs~\cite{adam_65} and which are also at the heart of the
approach of Goldstein for the relaxation dynamics of glass-forming
systems~\cite{goldstein_69}.

Finally we mention that we have found that upon a decrease of
temperature the sojourn time of the system within one MB increases
rapidly. This is in agreement with previous results in which the
concept of inherent structures was used~\cite{doliwa_03,vogel_04},
although here we have used a significantly simpler method to identify
the MB and which can notably also be used in real experiments such as
colloidal systems~\cite{kegel_00,weeks_00}. (We emphasize, however,
that we have obtained qualitatively the same results by considering
the inherent structures, although this approach is computationally much
more involved.)  On the other hand an increase of the temperature makes
that the structure of the MB is basically washed out and the ASD does
no longer show the pronounced peaks.

The present result indicate that the $\alpha-$relaxation is {\it not}
directly related to the presence of strings that are formed by a small
number of particles but instead is due to a cooperative rearrangement of a
substantial fraction of the particles. This cooperative rearrangement is
responsible for the transition between adjacent MBs and involves, at the
temperature considered, on the order of 40 particles.  During such
a MB-MB transition the majority of the particles {\it outside} this
CRR does not contribute significantly to the relaxation. Therefore one
needs on the order of 5-10 such transition events in order to complete an
$\alpha-$relaxation on the scale of the local neighborhood. Since these
transitions take only a few percent of the $\alpha-$relaxation time, one
thus can envision, at least at low temperatures, the $\alpha-$relaxation
as a sequence of rapid, localized, cooperative relaxation events in which
a substantial number of particles participate. This result is thus in
surprisingly good agreement with the picture put forward long time ago
by Adam and Gibbs and Goldstein.

Finally we mention that in view of the proposed connection between
the geometrical properties (distribution of barrier heights,
size of meta-basins,...) and the fragility of glass-forming
systems~\cite{debenedetti_01,angell_94}, it will be of great interest to do
the analysis presented here for a strong glass former. This will allow
to obtain a better understanding of the difference between strong and
fragile glass-formers~\cite{angell_85}.

GAA is research fellow of CONICET and RAM is research fellow of
CIC. Financial support from Fundaci\'on Antorchas, CONICET, UNS, SeCyT,
DYGLAGEMEM, and ECOS is gratefully acknowledged.


\vspace*{-7mm}

\begin{thebibliography}{10}

\bibitem{ediger_96}
M. D. Ediger, C. A. Angell, and S. R. Nagel, 
J. Phys. Chem. B {\bf 100}, 13200 (1996).

\bibitem{gotze_99}
W. G\"otze,
J. Phys.: Condens. Matter {\bf 10}, A1 (1999).

\bibitem{debenedetti_01}
P. G. Debenedetti and F. H. Stillinger,
Nature {\bf 410}, 259 (2001).

\bibitem{ngai_02}
K. L. Ngai (Ed.): \textit{Proceedings of the Forth International
Discussion Meeting on Relaxations in Complex Systems}, Non-Cryst. Solids
{\bf 307-310} (2002).

\bibitem{schmidt_91}
K. Schmidt-Rohr and H. W. Spiess,
Phys. Rev. Lett. {\bf 66}, 3020 (1991).

\bibitem{butler_91}
S. Butler and P. Harrowell,
J. Chem. Phys. {\bf 95}, 4454 (1991).

\bibitem{cicerone_95}
M. T. Cicerone, F. R. Blackburn, and M. D. Ediger, 
J. Chem.  Phys. {\bf 102}, 471 (1995).

\bibitem{kob_97}
W. Kob, C. Donati, S. J. Plimpton, P. H. Poole, and S. C. Glotzer, 
Phys. Rev. Lett. {\bf 79}, 2827 (1997).

\bibitem{kegel_00}
W. K. Kegel and A. van Blaaderen,
Science {\bf 287}, 290 (2000).

\bibitem{weeks_00}
E. R. Weeks, J. C. Crocker, A. C. Levitt, A. Schofield, and D. A. Weitz, 
Science {\bf 287}, 627 (2000).

\bibitem{richert_02}
R. Richert,
J. Phys.: Condens. Matter {\bf 14}, R703 (2002).

\bibitem{donati_98}
C. Donati, J. F. Douglas, W. Kob, S. J. Plimpton, P. H. Poole, and S. C. Glotzer,
Phys. Rev. Lett. {\bf 80}, 2338 (1998).

\bibitem{garrahan_02}
J. P. Garrahan and D. Chandler,
Phys. Rev. Lett. {\bf 89}, 035704 (2002).

\bibitem{angell_91}
C. A. Angell,
J. Non-Cryst. Solids {\bf 131}, 13 (1991).

\bibitem{debenedetti_01b}
P. G. Debenedetti, T. M. Truskett, and C. P. Lewis,
Adv. Chem. Eng. {\bf 28}, 21 (2001); 
F. Sciortino, J. Stat. Mech. (in press) (2005).

\bibitem{monthus_96}
C. Monthus and J.-P. Bouchaud,
J. Phys. A: Math. Gen. {\bf 29}, 3847 (1996).

\bibitem{doliwa_03}
B. Doliwa and A. Heuer,
Phys. Rev. E {\bf 67}, 031506 (2003).

\bibitem{vogel_04}
M. Vogel, B. Doliwa, A. Heuer, and S. C. Glotzer,
J. Chem. Phys. {\bf 120}, 4404 (2004).

\bibitem{kob_95}
W. Kob and H. C. Andersen,
Phys. Rev. E {\bf 51}, 4626 (1995).

\bibitem{ohmine_95}
I. Ohmine,
J. Phys. Chem. {\bf 99}, 6765 (1995).

\bibitem{footnote1}
In order to check for the presence of finite size effects we have
also repeated the presented analysis for a system with 1000 particles
and considered only a subbox in this extended system. The results
obtained were very similar to the ones presented here and therefore
finite size effects can be excluded. More details will be presented
elsewhere~\cite{appignanesi_05}.

\bibitem{appignanesi_05}
G. A. Appignanesi, J. A. Rodr\'{\i}guez Fris, R. A. Montani, and W. Kob,
to be published.

\bibitem{appignanesi_04}
G. A. Appignanesi and R. A. Montani,
J. Non-Cryst. Solids {\bf 337}, 109 (2004).

\bibitem{footnote2}
We mention that if this threshold is changed by a reasonable amount,
the number of mobile particles changes but the overall conclusions remain
nevertheless valid.

\bibitem{appignanesi_04b}
G. A. Appignanesi, M. A. Frechero, L. M. Alarcon, J. A. R. Fris, and  R. A. Montani,
Physica A {\bf 339}, 469 (2004).

\bibitem{adam_65}
G. Adam and J. H. Gibbs,
J. Chem. Phys. {\bf 43}, 139 (1965).

\bibitem{goldstein_69}
M. Goldstein, 
J. Chem. Phys. {\bf 51}, 3728 (1969).

\bibitem{angell_94}
C. A. Angell, P. H. Poole, and J. Shao,
Nuovo Cimento D {\bf 16}, 993 (1994).

\bibitem{angell_85}
C. A. Angell, p. 1 in: K. L. Ngai and G. B. Wright (Eds.) {\it Relaxation
in Complex Systems} (US Dept. Commerce, Springfield, 1985).

\end{thebibliography}
\end{document}